\documentclass[twocolumn,10pt,aps,pre,twoside,superscriptaddress,floatfix]{revtex4-1}

\pdfoutput=1

\usepackage{adjustbox}
\usepackage{amsmath}
\usepackage{amssymb}
\usepackage{graphicx}
\usepackage{dcolumn}
\usepackage{bm}
\usepackage{xcolor}
\usepackage[T1]{fontenc}

\def\dps{\displaystyle}

\def\p{\partial}
\def\epsilon{\varepsilon}

\def\theta{\vartheta}
\def\rho{\varrho}

\def\vec#1{\pmb{#1}}

\begin{document}


\title{The role of counterions in ionic liquid crystals}

\author{Hendrik Bartsch}
\affiliation
{
   Max-Planck-Institut f\"ur Intelligente Systeme,
   Heisenbergstr.\ 3,
   70569 Stuttgart,
   Germany
}
\affiliation
{
   Institut f\"ur Theoretische Physik IV,
   Universit\"at Stuttgart,
   Pfaffenwaldring 57,
   70569 Stuttgart,
   Germany
}
\author{Markus Bier}
\email{bier@is.mpg.de}
\affiliation
{
   Max-Planck-Institut f\"ur Intelligente Systeme, 
   Heisenbergstr.\ 3,
   70569 Stuttgart,
   Germany
}
\affiliation   
{
   Institut f\"ur Theoretische Physik IV,
   Universit\"at Stuttgart,
   Pfaffenwaldring 57,
   70569 Stuttgart,
   Germany
}
\affiliation
{
   Fakult\"{a}t Angewandte Natur- und Geisteswissenschaften,
   Hochschule f\"{u}r angewandte Wissenschaften\\
   W\"{u}rzburg-Schweinfurt,
   Ignaz-Sch\"{o}n-Str.\ 11,
   97421 Schweinfurt,
   Germany
}
\author{S.\ Dietrich}
\affiliation
{
   Max-Planck-Institut f\"ur Intelligente Systeme, 
   Heisenbergstr.\ 3,
   70569 Stuttgart,
   Germany
}
\affiliation
{
   Institut f\"ur Theoretische Physik IV,
   Universit\"at Stuttgart,
   Pfaffenwaldring 57,
   70569 Stuttgart,
   Germany
}

\date{23 October 2020}

\begin{abstract}
Previous theoretical studies of calamitic (i.e., rod-like) ionic liquid
crystals (ILCs) based on an effective one-species model led to indications of a
novel smectic-A phase with a layer spacing being much larger than the length of
the mesogenic (i.e., liquid-crystal forming) ions.
In order to rule out the possibility that this wide smectic-A phase is merely
an artifact caused by the one-species approximation, we investigate an
extension which accounts explicitly for cations and anions in ILCs.
Our present findings, obtained by grand canonical Monte Carlo simulations, show
that the phase transitions between the isotropic and the smectic-A phases of
the cation-anion system are in qualitative agreement with the effective
one-species model used in the preceding studies.
In particular, for ILCs with mesogenes (i.e., liquid-crystal forming species)
carrying charged sites at their tips, the wide smectic-A phase forms, at low
temperatures and within an intermediate density range, in between the isotropic
and a hexagonal crystal phase.
We find that in the ordinary smectic-A phase the spatial distribution of the
counterions of the mesogens is approximately uniform, whereas in the wide
smectic-A phase the small counterions accumulate in between the smectic layers.
Due to this phenomenology the wide smectic-A phase could be interesting for
applications which hinge on the presence of conductivity channels for mobile
ions.
\end{abstract}

\maketitle


\section{\label{sec:intro}Introduction}

Ionic liquid crystals (ILCs) are versatile materials which exhibit, on the one
hand, properties of ionic systems, such as the capability of charge transport,
and, on the other hand, they are able to form mesophases which is the
distinctive feature of liquid crystals~\cite{Binnemans2005,Goossens2016}.

Recently, several numerical studies~\cite{Kondrat2010,Saielli2017,Bartsch2017,
Bartsch2019} aimed {at} gaining insight {into} the link {between}
the underlying molecular features of ILCs and {their} resulting
properties, like {the} phase behavior or {the} formation of
nanostructures.
{The} complicated interplay of {anisotropy} and ionic molecular
properties renders {ILCs} {to be} challenging objects for theoretical
studies.
{Molecular dynamics} (MD) computer simulations can be performed with a
rather detailed description of the underlying molecules~\cite{Saielli2012}.
{Such} simulations can {provide} a rich and detailed picture of the
structures of the formed mesophases.
{But} the complexity of the underlying models makes it still
{difficult} to pinpoint basic characteristics of these systems {which
give} rise to the observed properties (or at least being essential for their
appearance).

In this regard, studying simplified models allows {one} to elucidate the
generic interplay of the key features of ILCs.
This might {provide} insight {into} elementary mechanisms on the
molecular level, which are necessary ingredients to the rich phenomenology of
ILC systems.
The natural drawback of simplistic models is that not all of the observable
features are captured.
At the same time, one is able to rule out minimal models, which are
{insufficient} to describe (or explain) particular phenomena of the
{object} of interest.
Thereby, it is possible to successively approach {and isolate} the minimal
physical requirements for the diverse phenomenology {occurring} for
complex systems such as ILCs.
{Moreover, this procedure yields} a generic understanding of the
individual microscopic mechanisms involved in the complex, mesoscopic and
macroscropic {materials} properties.

Typically the mesogenes {of} ILC systems are composed of long alkyl-chains
in combination with charged groups, like imidazolium rings~\cite{Binnemans2005}.
Due to the alkyl chains, ILC molecules exhibit a large aspect ratio, rendering
the molecules highly anisotropic in shape.
{However}, due to the flexibility of the molecules the formation of liquid
crystalline phases ({or} mesophases) is not self-evident.
Notwithstanding, in the spirit of the aforementioned simplified model
descriptions, within {the present} study we {refrain from focussing}
on the formation of orientationally ordered nanostructures, i.e., mesophases, by
\emph{flexible} molecules, which is typically driven by complex mechanisms like
microphase segregation~\cite{Binnemans2005,Goossens2016}.
Instead, {we consider} a coarse-grained model of \emph{rigid} particles, which
gives generically rise to the formation of mesophases ({in} particular
smectic phases) due to the underlying shape {of the molecules} and
{due to} the presence of van der Waals interactions.
{We} are aiming {at} elucidating the {influence} of charges on
the {formation of} liquid crystalline structures.
More specifically, we are concerned with the role of counterions {on} the
structure formation.
To this end, we {apply} a previously introduced model~\cite{Kondrat2010} of
ILCs, which {incorporates} only one of the ion species explicitly, while
the counterions are only considered as a homogeneous screening background.
Within this one-species ILC model an interesting phase behavior could be
observed which is very sensitive to the charge distribution within the
mesogenic ions~\cite{Bartsch2017}.

Inspired by ILCs {which are} based on positively charged imidazolium rings
attached to long alkyl chains~\cite{Goossens2016}, we adopt this notion and
refer to the mesogenic species as cations, while the spherical
\emph{counterions} are referred to as anions.

This study is structured as follows: In Sec.~\ref{sec:model} the model
{used} to describe the ILC molecules is {presented}, and additional
information {concerning} the methods to analyze the simulation data is
given.
In Sec.~\ref{sec:results} our present results, obtained by {grand
canonical} Monte Carlo (MC) simulations, are shown and discussed, followed by
our conclusions in Sec.~\ref{sec:conclusions}.


\section{\label{sec:model}Model and methods}

\subsection{\label{sec:model:pots}Pair interactions}

For common examples of ionic liquid crystals (ILCs), the cations ($\oplus$) are
characterized by a molecular structure exhibiting charged groups, e.g., in the
form of imidazolium rings, attached to rather long alkyl-chains, {whereas}
the anions ($\ominus$) are typically much smaller in size, e.g., in the form of
iodide ($I^{-}$)~\cite{Binnemans2005,Goossens2016}.
While the charged compounds introduce ionic properties, it is the presence of
the alkyl-chains {as part of} the cations which leads to the formation of
liquid-crystalline phases, so-called \emph{mesophases}, within ILCs.
Nonetheless, the occurrence of mesophases within ILCs, such as smectic phases,
is a highly non-trivial matter, because the internal flexibility of the
alkyl-chains hinders the formation of {layered} structures.
However, the interplay of the hydrophilic charged groups and the lipophilic
alkyl-chains stabilizes smectic structures via \emph{microphase segregation} of
the two molecular compounds~\cite{Binnemans2005}.

{In line with} the scope of {the present} study, we reduce the degree
of intra-molecular complexity by considering a coarse-grained description of
{ILCs}, in which the cations ($\oplus$) are represented by rigid
ellipsoids of length $L_\oplus$ and width $R_\oplus$, while the anions
($\ominus$) are spherical particles of diameter $R_\ominus<R_\oplus$.
Although such an approach does not allow {one} to study the underlying
mechanisms leading to the formation of smectic layers within the above
{described} {actual ILCs}, here we are interested in the dependence of the
smectic structures on the location of the charged groups.
Thus, the model parameters (see below) are tuned such that, within the
simplistic model, smectic phases are formed.
{We shall} analyze how these {layered} structures depend on the
intra-molecular location of the charges.

To this end, the pair interactions {used here consist} of
\begin{itemize}
\item[(i)]
 a hard-core contribution $\phi_{ij}^\text{hc}$,
 
\item[(ii)]
 an attractive energy contribution $\phi_{ij}^\text{vdW}$, accounting for
 {van der Waals} forces , and
 
\item[(iii)]
 the electrostatic interaction $\phi_{ij}^\text{es}$ due to the presence of
 charges.
\end{itemize}

Thus, the total pair potential $\phi_{ij}$ between two particles $i$ and $j$,
where $i,j\in\{\oplus,\ominus\}$, reads
\begin{align}
 \phi_{ij}=\phi^\text{hc}_{ij}+\phi^\text{vdW}_{ij}+\phi^\text{es}_{ij}.
 \label{eq:Tot_Pot}
\end{align}

As mentioned above, the cations are rigid prolate ellipsoids of
length-to-breadth ratio $L_\oplus/R_\oplus$.
Thus, the orientation of a cation is fully described by the direction
$\vec{\omega}$ of its long axis.
The total interaction potential between a pair of cations, i.e.,
Eq.~\eqref{eq:Tot_Pot} with $i=j=\oplus$, {is of} the following form:
\begin{widetext}
\begin{align}
 \phi_{\oplus\oplus}(\vec{r},\vec{\omega_1},\vec{\omega_2})= 
 \begin{cases}
  \infty
  &, 
  |\vec{r}| < R_\oplus\sigma(\vec{\hat r},\vec{\omega_1},\vec{\omega_2}), \\
  \phi^\text{GB}(\vec{r},\vec{\omega_1},\vec{\omega_2})+
  \phi_{\oplus\oplus}^\text{es}(\vec{r},\vec{\omega_1},\vec{\omega_2})
  &,
  |\vec{r}| \geq R_\oplus\sigma(\vec{\hat r},\vec{\omega_1},\vec{\omega_2}),
 \end{cases}
\label{eq:Cat_Cat_Pot}
\end{align}
\end{widetext}
where $\vec{r}$ denotes the center-to-center distance vector between two cations
with orientations $\vec{\omega_1}$ and $\vec{\omega_2}$.

The contact distance {between} two cations
\begin{align}
 \begin{split}
 R_\oplus\sigma(\vec{\hat r},\vec{\omega_1},\vec{\omega_2})
 =R_\oplus\biggl[1-\frac{\chi}{2}&\biggl(
   \frac{(\vec{\hat r}\cdot(\vec{\omega_1}+\vec{\omega_2}))^2}
        {1+\chi\,\vec{\omega_1}\cdot\vec{\omega_2}}\\
 +&\frac{(\vec{\hat r}\cdot(\vec{\omega_1}-\vec{\omega_2}))^2}
        {1-\chi\,\vec{\omega_1}\cdot\vec{\omega_2}}\biggr)\biggr]^{-1/2}
 \end{split}       
\label{eq:Cat_Cat_ContDis}
\end{align}
depends on the orientations of both cations and on the direction of the
center-to-center distance vector, expressed by the unit vector
$\vec{\hat r}:=\vec{r}/|\vec{r}|$.
In Eq.~\eqref{eq:Cat_Cat_Pot}, the contributions beyond the hard-core repulsion
at contact, i.e., for $|\vec{r}|\geq R_\oplus\sigma$, are subdivided into two
parts.
The attractive interactions $\phi_{\oplus\oplus}^\text{att}$, due to
{short-ranged} van der Waals forces between the cations,
{are} modeled by the Gay-Berne potential $\phi^\text{GB}(\vec{r},
\vec{\omega_1},\vec{\omega_2})=\phi_{\oplus\oplus}^\text{att}$~\cite{Berne1972,
Gay_Berne1981}.
$\phi^\text{GB}$ is a modification of the Lennard-Jones pair potential designed
for ellipsoidal particles:
\begin{widetext}
\begin{align}
  \phi_\text{GB}(\vec{r},\vec{\omega_1},\vec{\omega_2})=
  4\,\epsilon(\vec{\hat r},\vec{\omega_1},\vec{\omega_2})
 \biggl[\bigl(1+|\vec{r}|/R_\oplus-
  \sigma(\vec{\hat r},\vec{\omega_1},\vec{\omega_2})\bigr)^{-12} 
 -\bigl(1+|\vec{r}|/R_\oplus-
  \sigma(\vec{\hat r},\vec{\omega_1},\vec{\omega_2})\bigr)^{-6}\biggr]
\label{eq:GB_Pot}
\end{align}
\end{widetext}
with the anisotropic interaction strength
\begin{align}
 \begin{split}
 \epsilon(\vec{\hat r},\vec{\omega_1},\vec{\omega_2})&=\epsilon_{\oplus\oplus}
 \bigl(1-(\chi\,\vec{\omega_1}\cdot\vec{\omega_2})^2\bigr)^{-1/2}\\
 \times\biggl[1-\frac{\chi'}{2}\biggl(&
 \frac{(\vec{\hat r}\cdot(\vec{\omega_1}+\vec{\omega_2}))^2}
      {1+\chi'\,\vec{\omega_1}\cdot\vec{\omega_2}}+
 \frac{(\vec{\hat r}\cdot(\vec{\omega_1}-\vec{\omega_2}))^2}
      {1-\chi'\,\vec{\omega_1}\cdot\vec{\omega_2}}\biggr)\biggr]^{2}.
 \end{split}
\label{eq:GB_Pot_Eps}
\end{align}

{Both the} contact distance $R_\oplus\sigma$, i.e.,
Eq.~\eqref{eq:Cat_Cat_ContDis}, and the direction- and orientation-dependent
interaction strength {$\epsilon(\vec{\hat r},\vec{\omega_1},
\vec{\omega_2})$}, i.e., Eq.~\eqref{eq:GB_Pot_Eps}, {depend} on the cation
length-to-breadth ratio $L_\oplus/R_\oplus$ via $\chi=((L_\oplus/R_\oplus)^2-1)
/((L_\oplus/R_\oplus)^2+1)$.
Additionally, $\epsilon(\vec{\hat r},\vec{\omega_1},\vec{\omega_2})$ can be
tuned via $\chi'=((\epsilon_{R_\oplus}/\epsilon_{L_\oplus})^{1/2}-1)/
((\epsilon_{R_\oplus}/\epsilon_{L_\oplus})^{1/2}+1)$, where
$\epsilon_{R_\oplus}/\epsilon_{L_\oplus}$ is called the anisotropy parameter,
defined {as the ratio of the depth $\epsilon_{R_\oplus}$ of the Gay-Berne
potential minimum for parallel particles positioned side by side, i.e., with
$\vec{\hat r}\cdot\vec{\omega_1}=\vec{\hat r}\cdot\vec{\omega_2}=0$, and of
the depth $\epsilon_{L_\oplus}$ of the Gay-Berne potential minimum for parallel
particles positioned end to end, i.e., with $\vec{\hat r}\cdot\vec{\omega_1}=
\vec{\hat r}\cdot\vec{\omega_2}=1$.}
The length-to-breadth ratio $L_\oplus/R_\oplus$ and the anisotropy parameter
$\epsilon_{R_\oplus}/\epsilon_{L_\oplus}$ specify the molecular properties like
the shape and the chemical structure of the underlying cation molecules
within the present coarse-grained model.
As we aim for comparing the present 
findings with those of {our} previous studies using a one-species model
--- comprising only the ellipsoidal cations while the anions were incorporated
implicitly as a homogeneous screening background --- we choose the following
set of parameters for the Gay-Berne potential, which allows for such a
comparison with Ref.~\cite{Bartsch2017}:
\begin{align}
 L_\oplus/R_\oplus=4
 \hspace{15pt}\text{and}\hspace{15pt}
 \epsilon_{R_\oplus}/\epsilon_{L_\oplus}=3.
 \notag
\end{align}

It is worth mentioning, that in the case of spherical cations, i.e., for
$L_\oplus=R_\oplus$, Eq.~\eqref{eq:GB_Pot} reduces to the (isotropic)
Lennard-Jones potential iff $\epsilon_{R_\oplus}/\epsilon_{L_\oplus}=1$,
because {in that case} $\sigma(\vec{\hat r},\vec{\omega_1},\vec{\omega_2})
=1$ and $\epsilon(\vec{\hat r},\vec{\omega_1},\vec{\omega_2})=\epsilon_{\oplus
\oplus}$.
{The relations} $L_\oplus=R_\oplus$ and $\epsilon_{R_\oplus}\neq
\epsilon_{L_\oplus}$ describe molecules of rather spherical shape,
{which}, however, due to their internal chemical structure exhibit
non-spherical {van der Waals} interactions.

The remaining contribution $\phi_{\oplus\oplus}^\text{es}$ in
Eq.~\eqref{eq:Cat_Cat_Pot} is the \emph{e}lectro\emph{s}tatic repulsion between
the cations.
{Since for the present study} the electrostatic interactions are of
particular interest, their implementation is discussed in detail in the next
section.
At this point, we only point out that the charge sites are located symmetrically
{at} a distance $D$ from the cation center along the long axis.
Thus, $\phi_{\oplus\oplus}^\text{es}(\vec{r},\vec{\omega_1},\vec{\omega_2})$
depends not only on the distance $r=|\vec{r}|$ between the centers of two
cations, but also on the orientations $\vec{\omega_1}$ and $\vec{\omega_2}$, as
well as on the relative direction of the centers of the cations.

Anions are modeled as hard spheres of diameter $R_\ominus$ with a negative
charge site in {their} center.
To account for the omnipresent attractive {van der Waals} forces,
typically the Lennard-Jones potential is {used} to mimic these
dispersion-induced interactions between spherical particles.
However, here, we are neglecting any contributions arising from dispersion
forces between anions, i.e., $\phi_{\ominus\ominus}^\text{vdW}=0$, because we
focus on ionic effects originating from the much stronger electrostatic
interaction. {In} particular {we focus} on the influence of the anion
distribution on the liquid-crystalline structure of the cations.
Thus, the pair potential $\phi_{\ominus\ominus}(r)$ between two anions
separated by distance $r$ reads
\begin{align}
 \phi_{\ominus\ominus}(r)=
 \begin{cases}
  \infty  &, r < R_\ominus, \\ 
  \phi_{\ominus\ominus}^\text{es}(r)
  &,
  r \geq R_\ominus\ .
 \end{cases}
\label{eq:Anion_Anion_Pot}
\end{align}

$\phi_{\ominus\ominus}^\text{es}$ is the residual anion-anion electrostatic
interaction.
Due to the spherical shape of the anions it is an isotropic function and
depends only on the distance $r$. 
As mentioned above, further details of the electrostatic interactions are
provided in the next section.

The remaining anion-cation interaction potential
$\phi_{\oplus\ominus}=\phi_{\ominus\oplus}$ is defined as
\begin{align}
 \phi_{\oplus\ominus}(\vec{r},\vec{\omega})= 
 \begin{cases}
  \infty
  &, 
  r<R_{\oplus\ominus}\delta(\vec{\hat r}\cdot\vec{\omega}), \\
  \phi_{\oplus\ominus}^\text{es}(\vec{r},\vec{\omega})
  &,
  r\geq R_{\oplus\ominus}\delta(\vec{\hat r}\cdot\vec{\omega})\ .
 \end{cases}
\label{eq:Cat_Anion_Pot}
\end{align}
The contact distance $R_{\oplus\ominus}\delta(\vec{\hat r}\cdot\vec{\omega})$
between an ellipsoidal cation $\oplus$ and a spherical anion $\ominus$
can be expressed as the product of the minimal contact distance
$R_{\oplus\ominus}:=(R_\oplus+R_\ominus)/2$
(obtained for $\vec{\hat r}\cdot\vec{\omega}=0$)
multiplied by the elliptical scaling function
\begin{align}
 \delta(\vec{\hat r}\cdot\vec{\omega})=
 \big[1-\chi_{\oplus\ominus}(\vec{\hat r}\cdot\vec{\omega})^2\big]^{-1/2},
\label{eq:Cat_Anion_ContDis}
\end{align}
where $\chi_{\oplus\ominus}:=1-R_{\oplus\ominus}^2/L_{\oplus\ominus}^2$ such
that the contact distance $R_{\oplus\ominus}\delta(\vec{\hat r}\cdot
\vec{\omega})$ reaches its maximum value $L_{\oplus\ominus}:=(L_\oplus+
R_\ominus)/2$ for $\vec{\hat r}\cdot\vec{\omega}=1$.

{We do not consider} contributions due to {van der Waals} forces
between cations and anions, i.e., $\phi_{\oplus\ominus}^\text{vdW}=0$.
{These contributions} might be necessary in order to describe
quantitatively reliably a specific type of ionic liquid crystal system.
{Here, however,} we are not interested in such a quantitative analysis,
but we are rather aiming at a general understanding of the mechanisms
leading to structures and distinct phases in ILCs.
{In particular, we want to understand} how {a possibly non-uniform
counterion distribution} affects the phase behavior {which} is observed
within the effective one-species model used in Ref.~\cite{Bartsch2017}.
In practice, this means that the effectively treated electrostatic interaction
is altered such that the valency dependence of the Coulomb potential is now
explicitly incorporated.
{Since} this is a key issue of the present {study}, the
implementation of the electrostatic interactions among all particles is
explicitly given in the following section.

Finally, we point out that the remaining independent parameters $R_\oplus$
(which denotes the cation width, see Eq.~\eqref{eq:Cat_Cat_ContDis}) and the
cation-cation interaction constant $\epsilon_{\oplus\oplus}$ (see
Eq.~\eqref{eq:GB_Pot_Eps}), are chosen as the length and the energy scale of
the system.


\subsection{\label{sec:model:elecstat}Electrostatic energy contributions}

Within the present model, both cations ($\oplus$) and anions ($\ominus$) carry
point-like charge sites.
While each anion carries a single charge site in its center, cations exhibit
two distinct charge sites, located {at} a distance $D$ from their
geometrical center.
Thus, the electrostatic interactions among all types of particles are given by
\begin{align} 
 \phi_{\ominus\ominus}^\text{es}
 &= 4\,\gamma\,\tilde\phi(r),
 \label{eq:Anion_Anion_Pot_es}\\ 
 \phi_{\oplus \oplus }^\text{es}
 &=\gamma\,\bigl(
   \tilde\phi(|\vec{r}+D(\vec{\omega_1}+\vec{\omega_2})|)\notag\\
 &\ \ \ \ +\tilde\phi(|\vec{r}+D(\vec{\omega_1}-\vec{\omega_2})|)\notag\\
 &\ \ \ \ +\tilde\phi(|\vec{r}-D(\vec{\omega_1}+\vec{\omega_2})|)\notag\\
 &\ \ \ \ + \tilde\phi(|\vec{r}-D(\vec{\omega_1}-\vec{\omega_2})|)\bigr),
 \label{eq:Cat_Cat_Pot_es}
\end{align}
{and}
\begin{align}
 \phi_{\oplus \ominus}^\text{es}
 &= -2\,\gamma\bigl(
   \tilde\phi(|\vec{r}+D\,\vec{\omega}|)+
   \tilde\phi(|\vec{r}-D\,\vec{\omega}|)\bigr)
 \label{eq:Cat_Anion_Pot_es}
\end{align}
with {the} electrostatic interaction strength $\gamma=q^2/(4\pi
\epsilon_0)$, where {$\epsilon_0$} denotes the vacuum permittivity,
and {$q>0$} is the charge of a single site of the cations.
The {factors of} $4$ in Eq.~\eqref{eq:Anion_Anion_Pot_es} and {of} $2$
in Eq.~\eqref{eq:Cat_Anion_Pot_es} occur, because the negative charge site in
the center of an anion {has to} be twice as strong as a single cation
charge site, such that the valency is the same for cations and anions.
Thus, in order to guarantee \emph{global charge neutrality}, the system
contains the same number of cations and anions (this issue will be discussed in
detail in the next section).
We note, that the factor {of} $4$ is also recovered in
Eqs.~\eqref{eq:Cat_Cat_Pot_es} and \eqref{eq:Cat_Anion_Pot_es} for $D=0$.
In Eqs.~\eqref{eq:Anion_Anion_Pot_es} and \eqref{eq:Cat_Cat_Pot_es} the
electrostatic energy contributions $\phi_{\ominus\ominus}^\text{es}\geq0$ and
$\phi_{\oplus\oplus}^\text{es}\geq0$ are repulsive, while in
Eq.~\eqref{eq:Cat_Anion_Pot_es} the negative sign indicates the electrostatic
attraction of cations and anions, i.e., $\phi_{\oplus\ominus}^\text{es}\leq0$.

For point-like charges $q$ in $d=3$ dimensions, the electrostatic interaction
potential decays as {the inverse of the distance}, i.e., $\tilde\phi(r)=
1/r$ {(see Eq.~\eqref{eq:Anion_Anion_Pot_es})}.
Thus, it is {long-ranged} and this property is well-known to lead to a
wide range of {peculiarities} of ionic systems~\cite{Hansen1986}.
{In the present context it is important to note}, that, in order to
accurately account for the {long-ranged} character {of the
interactions}, in computer simulations (MC or MD) of bulk systems, {based
on} \emph{periodic boundary conditions}, one {has} to {resort to}
sophisticated methods like the \emph{Ewald summation}~\cite{Ewald1921}.
The Ewald summation splits the full electrostatic contribution to the total
energy of a given configuration into a {short-ranged} and a
{long-ranged} part by expanding the actual charge density by a set of
Gaussian screening charge clouds.
While the first contribution is a sum over {short-ranged} interaction
potentials and can be calculated in real-space, the second contribution
contains the long-ranged {part} which can be calculated by
Fourier transformation {by expoiting} the periodic boundary conditions.
A different perspective on this method is, that the Ewald summation separates
the electrostatic energy into two contributions, such that the first one
expresses the valency dependence of the Coulomb interaction, while the second
one is determined by the {long-ranged part}.

The motivation {for} the present {study} {is} to
analyze the effects incorporated {due to} accounting for \emph{both} ion
species {and to {compare} them with those occurring within} the effective 
one-species model {which} has been used previously for studying the bulk
phase behavior of ILCs~\cite{Kondrat2010,Bartsch2017}.
{First}, it is interesting to study a system of cations and anions
interacting via {short-ranged} potentials and analyze how the valency
dependence affects the previous results.
{In a second step} the full Ewald summation allows {one} to
investigate the {relevance} of the {long-ranged} character.
In this way one can gain insight into the influence of both 
these two fundamental properties of the Coulomb interaction on the phase
behavior of ILCs.

The interaction potential, resembling the {short-ranged} contributions to
the total electrostatic energy contribution, {is} described by a Yukawa
potential
\begin{align}
 \tilde\phi(r):=\frac{\exp\bigl(-r/\lambda\bigr)}{r}
\label{eq:Yukawa}
\end{align}
{(see Eq.~\eqref{eq:Anion_Anion_Pot_es})} with decay length $\lambda=5
R_\oplus$ (Eqs.~\eqref{eq:Anion_Anion_Pot_es}-\eqref{eq:Cat_Anion_Pot_es}).
While Eq.~\eqref{eq:Yukawa} exhibits the same functional form {as the one}
used in Refs.~\cite{Kondrat2010,Bartsch2017}, it is important to
{emphasize} that the decisive {new aspect of the present study} is
the actual presence of counterions, i.e., positive (repulsive) \emph{and}
negative (attractive) electrostatic energy contributions.
{The decay length} $\lambda=5R_\oplus$ has been chosen {such as} to
match the parameters {of the interaction potential} in
Ref.~\cite{Bartsch2017}.
In addition, for the given cation length $L_\oplus=4R_\oplus$ and the
anion diameter $R_\ominus<R_\oplus$, the chosen decay length $\lambda=
5R_\oplus$ is larger than the particle sizes, such that Eq.~\eqref{eq:Yukawa}
corresponds to a weak artificial screening of the pure Coulomb potential $1/r$.
A previous study~\cite{Bartsch2015} suggests that the valency dependence rather
than the long-ranged character of the electrostatic interaction is decisive
for the phase behavior of ionic fluids, i.e., the weak artificial screening
in Eq.~\eqref{eq:Yukawa} is expected to give rise to at most some
quantitative consequences.
Moreover, in Ref.~\cite{Stenqvist2019}, it has been shown recently that there
are plenty of alternatives to the functional form of Eq.~\eqref{eq:Yukawa},
which serve to describe the structure of actual ionic systems remarkably well.
Thus we expect that incorporating the full Coulomb interaction via the
Ewald method only leads to a quantitative change {of} the phase behavior,
{such as a shift} of the phase transitions in {the}
temperature-density {plane, which, however}, does not significantly alter
the {occurrence of the} phases {or} their {structural properties}
on a qualitative level.

\begin{figure}
\includegraphics[width=8cm]{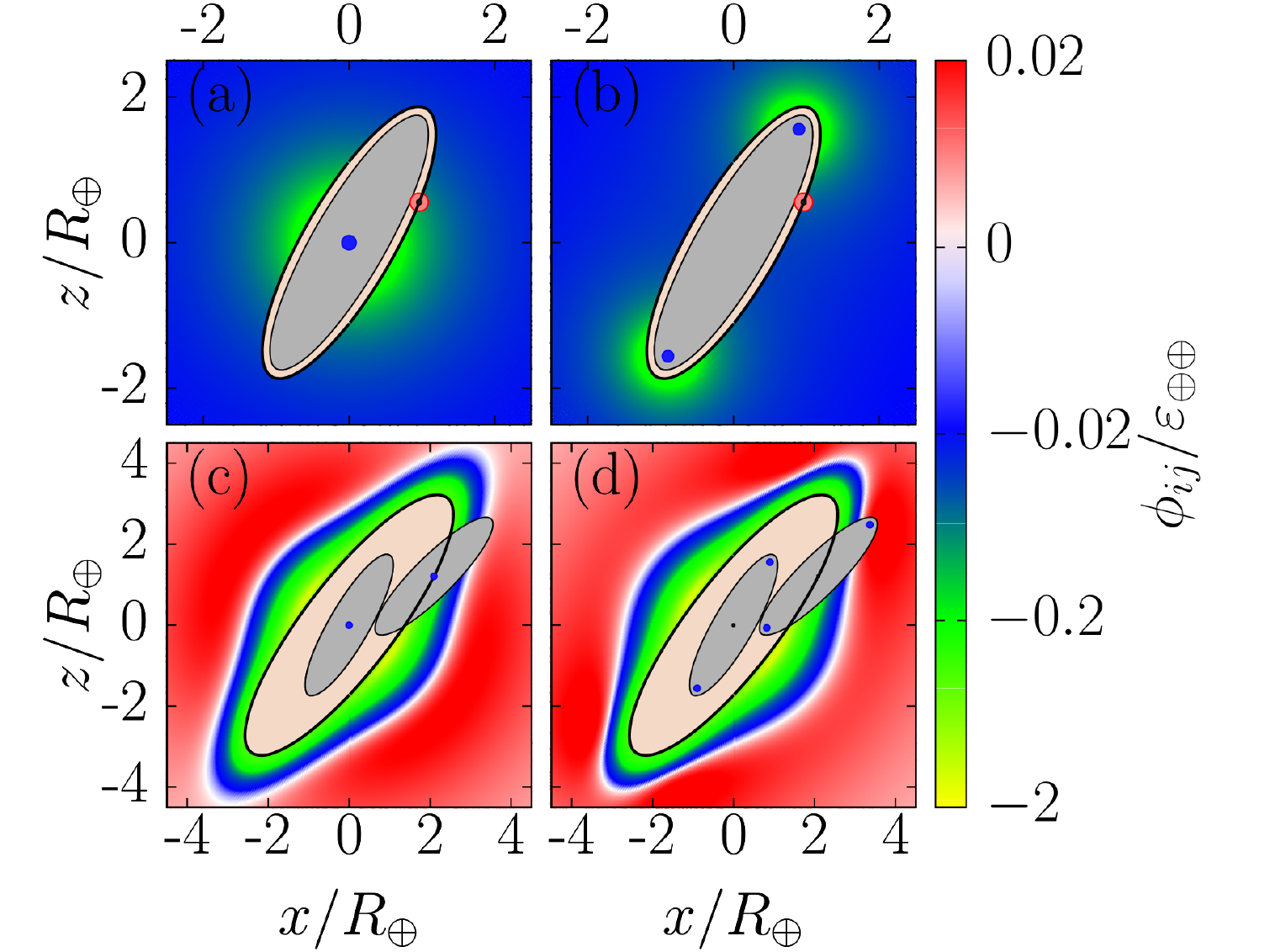}
\caption{Contour plots of the total interaction potentials
 $\phi_{ij}=\phi_{\oplus\ominus}$ (Eq.~\eqref{eq:Cat_Anion_Pot}), acting between
 an ellipsoidal cation ($\oplus$, gray-colored) and a {small} spherical anion
 ($\ominus$, red-colored, located at $x/R_\oplus\approx1,z/R_\oplus\approx
 0.6$), are shown in panels (a) and (b) for $D=0$ {and} for
 $D/R_\oplus=1.8$, respectively; the orientation $\vec{\omega}$ of the
 cation is fixed.
 The black solid line marks the rim of the small bright excluded volume
 between the cation and the anion.
 Beyond the contact distance the pair interaction is purely electrostatic and
 attractive, with {the} attraction {strongest} close to the center
 for $D=0$ (a) and close to the tips for $D/R_\oplus=1.8$ (b), respectively.
 Panels (c) and (d) provide the cation-cation interaction potentials
 $\phi_{ij}=\phi_{\oplus\oplus}$ for the same two types of charge distributions
 of the cations {as in (a) and (b)}.
 At very short distances the attractive Gay-Berne interaction is dominant,
 while the electrostatic repulsion is dominant {at large distances}.
 {In (c) and (d) the} attraction is strongest for a parallel orientation
 of the two cations {with} a side-to-side positioning.
 The location of strongest repulsion depends on the position of
 the charges.
 The bright excluded volumes in (c) and (d) are much larger than in (a)
 and (b).
 Here and in the simulations we have used the parameter values
 $L_\oplus/R_\oplus=4$,
 $\epsilon_{R_\oplus}/\epsilon_{L_\oplus}=3$, $R_\ominus/R_\oplus=1/4$,
 $\gamma/(\epsilon_{\oplus\oplus}R_\oplus)=0.045$, and $\lambda/R_\oplus=5$.}
\label{fig:1}
\end{figure}

In Fig.~\ref{fig:1} the full potentials for the cation-anion and the
cation-cation interactions are presented.
While panels (a) and (b) show the full cation-anion interaction for $D=0$ and
$D/R_\oplus=1.8$, respectively, in panels (c) and (d) the cation-cation
interaction potential is illustrated for the same types of charge distributions
of the cations {as in (a) and (b), respectively}.
For the considered particle sizes, i.e., $L_\oplus/R_\oplus=4$ and $R_\ominus/
R_\oplus=1/4$, the excluded volume (illustrated by the {beige} area with
{a} solid black rim) {between} cations and anions is much smaller compared
to the excluded volume between two cations.
{If $D=0$}, the electrostatic interaction is strongest for a side-by-side
position of the two considered particles, while {for $D/R_\oplus=1.8$} it
is strongest at the tips.
Moreover, the additional electrostatic repulsion among cations leads to a
narrowing of the most attractive region (which stems from the attractive
Gay-Berne interaction) at close distances for a side-by-side configuration of
{the} cations, if the cation charges are located {at the center of the
molecule}.


\subsection{\label{sec:methods:pdf}Pair distribution functions}

{In the course of} the simulations the structure of the fluid is analyzed
via pair distribution functions.
For simulations of bulk systems, the pair distribution functions 
$g_{ij}(\vec{r},\vec{r'}):=\rho_{ij}(\vec{r}|\vec{r}')/\rho_j$ can be defined
as the ratio of the \emph{conditional} density $\rho_{ij}(\vec{r}|\vec{r}')$ of
particles of species $i$ at position $\vec{r}$, provided a particle of species
$j$ is located at $\vec{r'}$, and of the {constant} mean density $\rho_j$
in the simulation box~\cite{Hansen1986,Allen1989}.

As we are mainly interested in observing (smectic) layer structures, we monitor
the pair distribution in the direction of the layer normal $\vec{\hat n}$, as
well as the {particle} distribution within the layers, i.e., in directions
$\vec{r}_\perp$ {lateral} to the layer normal.
Parallel to the layer normal $\vec{\hat n}$, {the} statistics along the
simulation trajectories {invokes} all pairs of particles {at}
distances $n:=|(\vec{r}-\vec{r'})\cdot\vec{\hat n}|$:
\begin{align}
 g_{ij}^{||}(n):=\frac{\rho_{ij}^{||}(n)}{\rho_j}.
\label{eq:PairDistFunc_normal}
\end{align}

Additionally, for the planes perpendicular to $\vec{\hat n}$ --- associated
with the vector $\vec{r}_\perp:=\vec{r}-(\vec{r}\cdot\vec{\hat n})\vec{\hat n}$
--- we monitor the radial distribution of cation pairs via
\begin{align}
 g_{\oplus\oplus}^{(n)}(r_\perp):=
 \frac{\rho_{\oplus\oplus}^{(n)}(r_\perp)}{\rho_\oplus},
\label{eq:PairDistFunc_lateral}
\end{align}
where $n\in\{0,d\}$ refers to the plane for which $n=0$ {and} $n=d$,
{respectively}.
Thus $g_{\oplus\oplus}^{(0)}(r_\perp)$ monitors the (cation) pair distribution
in the $0$-th plane, i.e., the plane which contains the reference cation at
$\vec{r'}$, while $g_{\oplus\oplus}^{(d)}(r_\perp)$ monitors the distribution
of particles in the two neighboring layers, with respect to the cation at
$\vec{r'}$.

We add the following {remarks} concerning the {computation} of
Eqs.~\eqref{eq:PairDistFunc_normal} and \eqref{eq:PairDistFunc_lateral}:
\begin{itemize}
\item
 The direction of the layer normal $\vec{\hat n}$ is {determined} by
 calculating the director~\cite{DeGennes1974} of each configuration along the
 simulated trajectories.
 For the relevant cases, {which} are analyzed within the scope of
 {the present study}, the director and the layer normal point {into}
 the same direction, i.e., they are (almost) parallel.
 
\item
 The evaluation of the conditional densities $\rho_{ij}^{||}(n)$ and
 $\rho_{\oplus\oplus}^{(n)}(r_\perp)$ requires to count the number of particles
 {which} are a distance $n$ {and} a distance $r_\perp$,
 {respectively}, apart from the central reference particle.
 To this end, one considers {small but nonzero} volumina at $n$ and
 $r_\perp$, which are given by {straight} slices of width $\Delta n$ and
 by annuli of width $\Delta r_\perp$, respectively.
 The {straight} slices for calculating $\rho_{ij}^{||}(n)$ {extend}
 in lateral direction up to the boundaries of the simulation box (see the
 illustration of such a slice in Fig.~\ref{fig:2}(a)), while the annuli for
 calculating $\rho_{\oplus\oplus}^{(n)}(r_\perp)$ have an extent in the
 direction of the layer normal {from} $\Delta_{||}=1\times R_\oplus$ to
 $\Delta_{||}=2\times R_\oplus$.
 
\item
 While the volume of an {annulus} is given by $\Delta V^{\perp}=\pi
 \Delta_{||}\Delta r_\perp(\Delta r_\perp+2r_\perp)$, the volumina
 $\Delta V^{||}$ of the {straight} slices cannot be calculated
 straightforwardly, as they are cut off at the boundaries of the simulation box.
 However, by considering a reference configuration with a homogeneous and
 isotropic distribution of particles {in the simulation box}, the volumina
 of the {straight} slices can be approximated by $\Delta V^{||}\approx\bar
 N/\rho^\text{ref}$, where $\bar N$ denotes the number of particles counted
 within the considered slices for the isotropic and homogeneous reference
 configuration of mean density $\rho^\text{ref}$.
\end{itemize}


\section{\label{sec:results}Results}

Before presenting our results, we note that the {grand canonical} Monte
Carlo simulations were performed within cubic simulation boxes of volume
$V=L^3$, where $L:=15 R_\oplus$.
The cation breadth $R_\oplus$ is chosen as the unit {of} length.
Standard Metropolis importance sampling {has been
used}~\cite{Metropolis1953}, invoking the configurational acceptance function
$S(\zeta):=-\beta H(\zeta)+\beta\mu N(\zeta)-2\ln((N(\zeta)/2)!)$.
{Here}, $\beta H(\zeta)$ denotes the total (potential) energy of a given
configuration $\zeta$ ({see} Sec.~\ref{sec:model:pots} and
Eq.~\eqref{eq:Tot_Pot}) {in units of the} thermal energy {$k_BT=
\beta^{-1}$}, $\beta\mu$ is the chemical potential, and $N(\zeta)$ {is} the
total number of particles.
Since we are considering 1:1-ionic mixtures, {there is an equal} number of
cations and anions, i.e., $N_\oplus(\zeta)=N_\ominus(\zeta)=N(\zeta)/2$.

Since we are mainly concerned with (smectic) structures formed by the
mesogenic cations, {the number} density is {given} in terms of the
cation packing fraction $\eta:=\frac{\pi}{6}L_\oplus R_\oplus^2\langle N_\oplus
\rangle/L^3$, where $\frac{\pi}{6}L_\oplus R_\oplus^2$ denotes the cation
volume and {$\langle N_\oplus\rangle$} refers to the thermally averaged
total number of cations.
Temperature is measured in terms of the ratio of {the} thermal energy
{$k_BT$ and} the interaction strength $\epsilon_{\oplus\oplus}$ of the
Gay-Berne potential (Eq.~\eqref{eq:GB_Pot_Eps}), i.e., {$T^*:=k_BT/
\epsilon_{\oplus\oplus}$}.

If not stated otherwise, for the simulations we used the following model
parameters: $L_\oplus/R_\oplus=4$, $\epsilon_{R_\oplus}/\epsilon_{L_\oplus}=3$,
$R_\ominus/R_\oplus=1/4$, $\gamma/(\epsilon_{\oplus\oplus}R_\oplus)=0.045$,
{and} $\lambda/R_\oplus=5$.
Furthermore, for calculating the total energy, all pair interactions
(Eq.~\eqref{eq:Tot_Pot}) {have been truncated} beyond {the range}
$R_\text{cut}/R_\oplus=6$.

The phase diagrams displayed in Fig.~\ref{fig:4} are obtained by
performing simulations for numerous state points $(T^*,\beta\mu)$, whereby
each run is initialized with an \emph{isotropic} configuration.
By performing short additional simulation runs initialized with smectic-A and
crystalline configrations, it has been checked for all considered state points
that the simulation results do not depend on the initialization.
The phase transitions in Fig.~\ref{fig:4} are resolved with an accuracy of
$\Delta(\beta\mu)=0.1$ in terms of the chemical potential $\beta\mu$.
Taking into account the values of $\dps\frac{\p\eta}{\p(\beta\mu)}$ obtained
from the simulations, this accuracy is sufficient to resolve the white two-phase
regions in Fig.~\ref{fig:4} with an accuracy of $\dps\Delta\eta\approx
\frac{\p\eta}{\p(\beta\mu)}\,\Delta(\beta\mu)\leq0.01$ in terms of the packing
fraction $\eta$.


\subsection{\label{sec:results:struc_dep}Dependence of smectic structures
on the charge distribution {within the cations}}

\subsubsection{\label{sec:results:struc_dep:SAW}Formation of the phase $S_{AW}$}

\begin{figure}
\includegraphics[width=8cm]{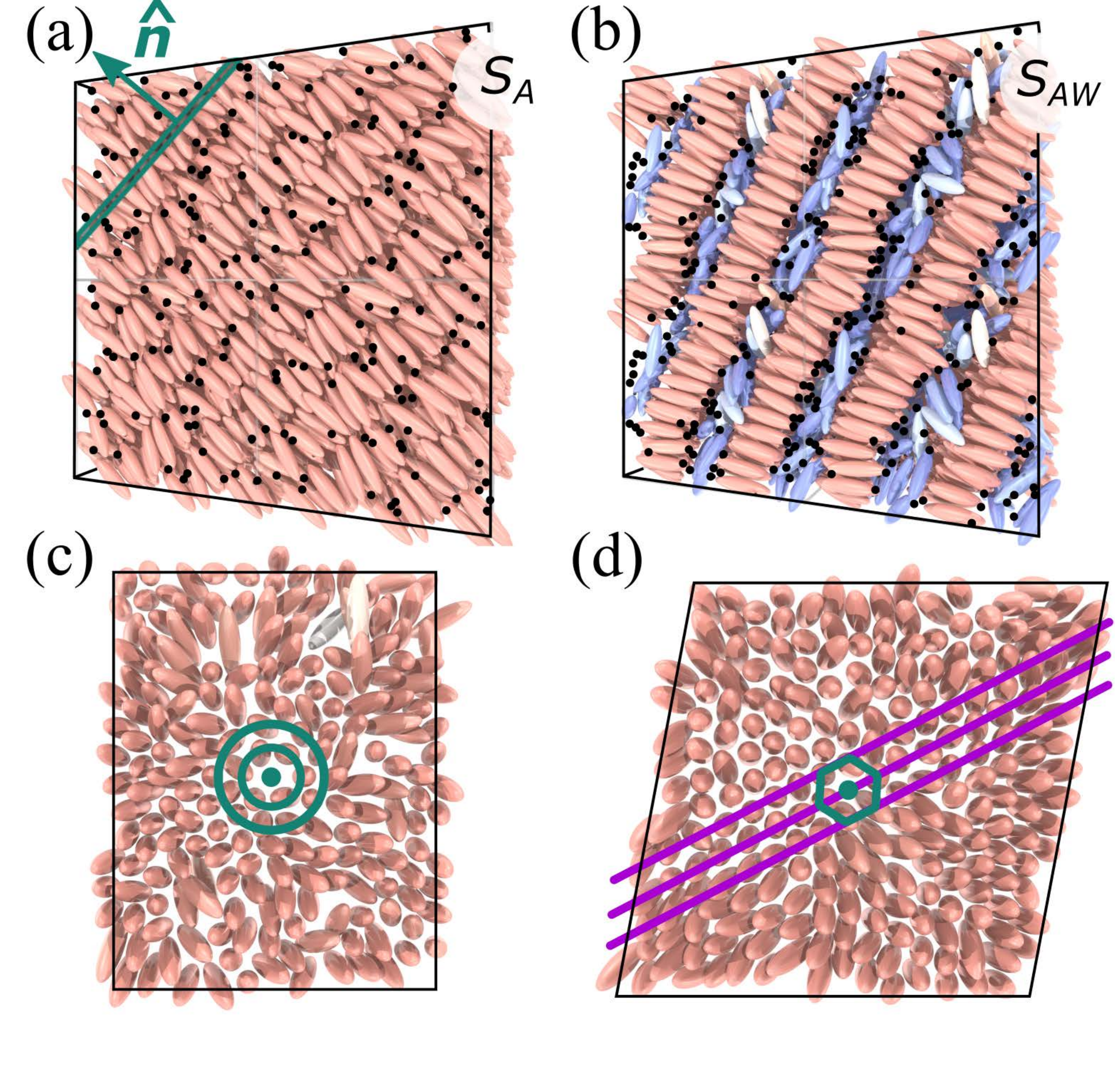}
\caption{In (a) and (b) we show snapshots of two configurations,
 belonging to the phases $S_A$ and $S_{AW}$, respectively.
 While for the phase $S_A$ a layer spacing of the size of the cation length
 $L_\oplus$ can be observed, and all cations tend to be aligned with the smectic
 layer normal $\vec{\hat n}$, in (b) the alternating layer structure of the
 phase $S_{AW}$ is clearly visible.
 In between the layers of cations, being well-aligned with the layer normal
 $\vec{\hat n}$ (red-colored ellipsoids), secondary layers are observed, in
 which the cations are preferentially perpendicular to the layer normal (blue
 ellipsoids).
 Due to the alternating layer structure the layer spacing is significantly
 increased.
In panels (a) and (b) the anions are depicted as small black dots,
 which, however, in order to increase visibility, are three times larger than
 the actual anions ($R_\ominus/R_\oplus= 1/4$).
 In (c) and (d) the intra-layer structures of the {phases} $S_A$ and
 $S_{AW}$, respectively, are shown.
 While in (c) {for the phase $S_A$} a fluid-like structure is observed,
 the {snapshot} of a main layer of the phase {$S_{AW}$} resembles a
 hexagonal structure (highlighted by the green hexagon and the thick violet
 lines in panel (d)).
 This observation is accompanied by a {higher} cation density within the
 main layers of the phase $S_{AW}$ as compared {with} the $S_A$ layers.
 Note, that the green slab in the upper left corner of panel (a)
 depicts a slice which is
 used to evaluate the pair distribution functions $g^{||}_{ij}(n)$ in the
 direction of the layer normal $\vec{\hat n}$ (green arrow).
 Similarly, in panel (c) the green concentric circles {indicate} the
 annulus for calculating the lateral pair distribution function
 $g^{(n)}_{\oplus\oplus}(r_\perp)$.}
\label{fig:2}
\end{figure}

\begin{figure*}
\includegraphics[width=16cm]{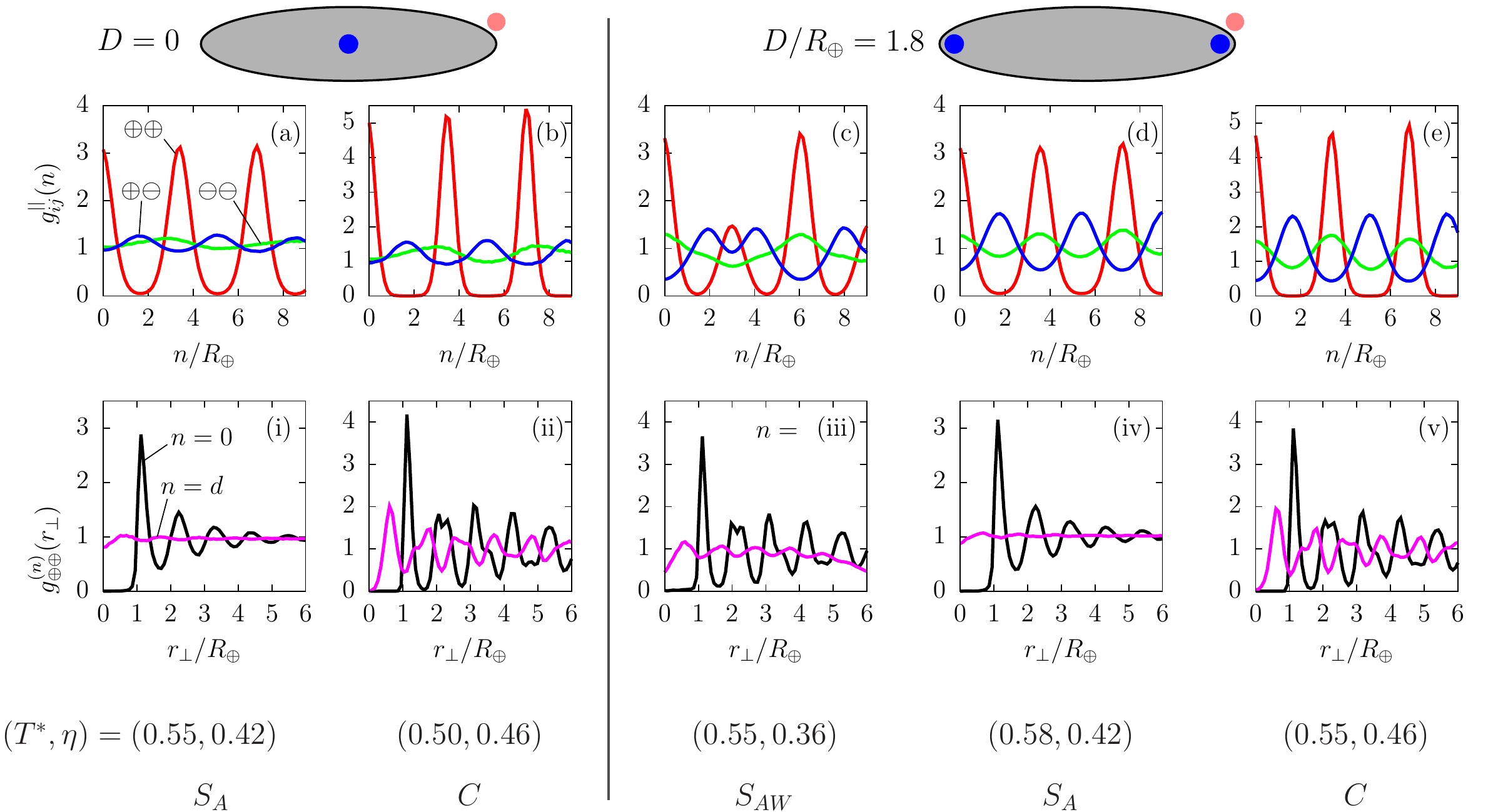}
\caption{Panels (a) and (b) show the pair distribution functions $g_{ij}^{||}
 (n)$ in the direction of the smectic layer normal for $D=0$, while panels
 (c)--(e) correspond to the case $D/R_\oplus=1.8$.
 In the second row, i.e., panels (i)-(v), the corresponding lateral pair
 distribution functions $g_{\oplus\oplus}^{(n)}(r_\perp)$ are plotted.
 (Analogously to the first row, panels (i) and (ii) refer to the case $D=0$ and
 panels (iii) to (v) to the case $D/R_\oplus=1.8$.)
 In panel (a) a layering of cations with layer spacing $d/R_\oplus\approx3.5
 \approx L_\oplus/R_\oplus$ {can be inferred} from $g_{\oplus\oplus}^{||}
 (n)$ (red curve).
 From the corresponding lateral pair distribution function
 $g_{\oplus\oplus}^{(0)}(r_\perp)$ (black solid line) in panel (i), a
 fluid-like structure within the layers can be inferred.
 Thus, an ordinary smectic-A phase ($S_A$) is formed.
 (The data shown in panels (a) and (i) correspond to the state point
 {$(T^*=0.55,\eta\approx0.42)$}, see Fig.~\ref{fig:4}.) 
 In panels (b) and (ii) a similar layer structure of cations with $d/R_\oplus
 \approx3.5$ is observed.
 {However,} {the corresponding state point
 $(T^*=0.5,\eta\approx0.46)$ is located at a lower temperature $T^*$ and
 at a higher density $\eta$}.
 The strong lateral correlations --- even among neighboring layers (see
 the magenta curve ``$n=d$'' in panel (ii)) --- {indicate} that this refers to
 a  hexagonal crystal $C$.
 Both structures for $D=0$ exhibit only a {weakly} inhomogeneous
 distribution of anions, i.e., $g_{\oplus\ominus}^{||}(n)$ and
 $g_{\ominus\ominus}^{||}(n)$ {exhibit} only minor variations {as
 function of the} distance $r_\perp$.
 Panels (c) and (iii) refer to the phase {$S_{AW}$} formed for
 $D/R_\oplus=1.8$ at low {temperatures} and intermediate densities
 {$(T^*=0.55, \eta\approx0.36)$}.
 The alternating layer structure of cations with significantly larger layer
 spacing $d/R_\oplus\approx6>L_\oplus/R_\oplus$ is apparent from $g_{\oplus
 \oplus}^{||}(n)$.
 Due to the enhanced density within the main cation layers (see the maxima of
 $g_{\oplus\oplus}^{||}(n)$ at $n=0$ and $n/R_\oplus\approx6$) in lateral
 directions a hexagonal structure can be observed, unlike the fluid-like
 {lateral} structure of the ordinary phase $S_A$.
 However, correlations between cations in neighboring layers are almost absent
 and thus the phase $S_{AW}$ is a genuine smectic phase and not a crystal.
 {We note} that the small variations in $g_{\oplus\oplus}^{(d)}(r_\perp)$
 are artifacts due to the periodic boundary conditions (see the
 {discussion in Sec.~\ref{sec:results:struc_dep} of the} main text) and
 the drop of $g_{\oplus\oplus}^{(d)}(r_\perp)$ at {large} distances
 $r_\perp>5$ is another artifact, due to insufficient statistics.
 Interestingly, for the phase $S_{AW}$ a considerable inhomogeneous
 distribution of anions (in normal direction) can be observed in panel (c).
 The anions prefer to be close to the locations of the cation charges in the
 main layers, e.g., at $n/R_\oplus\approx1.8$, as can be inferred from
 $g_{\oplus\ominus}^{||}(n)$.
 Also for $D/R_\oplus=1.8$ the ordinary phase $S_A$ (at higher temperature
 $T^*=0.58$ and $\eta\approx0.42$) and the hexagonal crystal $C$ (at $T^*=0.55$
 and $\eta\approx0.46$) can be observed (see panels (d) and (iv), respectively
 (e) and (v)).
 However, in contrast to the findings for the corresponding phases for $D=0$,
 for $D/R_\oplus=1.8$ the anions {exhibit a considerably} inhomogeneous
 spatial distribution.
 They prefer to be located in between the cation layers, similar to the
  phase {$S_{AW}$}.}
\label{fig:3}
\end{figure*}

The present model (see Sec.~\ref{sec:model}) can be understood as an extension
of the effective one-species model of {ILCs} used in previous theoretical
studies~\cite{Kondrat2010,Bartsch2015,Bartsch2017}, such that the present,
extended model accounts for {the} explicit presence of \emph{both} ionic
species.
The present approach allows us to study explicitly the effect of incorporating
the valency dependence of the Coulomb interaction on top of the previously
studied one-species model.
We note, that yet it is necessary to introduce hard-core interactions between
the cations and anions, as well as among the anions, in order to avoid
divergences in the electrostatic interactions (caused by mutual penetration).

One of the most striking findings within the effective one-species model
in Ref.~\cite{Bartsch2017} {is} the sensitive dependence of the occurring
smectic structures on the location of charges {within} the ellipsoidal
cations.
Therefore, we {first} analyze how the presence of counterions affects this
dependence.

We start our analysis at fixed temperature $T^*=0.55$ and discuss the structures
which are formed at sufficiently {high} densities, such that the isotropic
phase becomes {thermodynamically} unstable (or at least metastable) with
respect to smectic phases.
{We note} that, for the considered parameters, no nematic phase is
observed in the density regime between the isotropic and the smectic phase.
(For larger values of the cation length-to-breadth ratio $L_\oplus/R_\oplus$
this might, {however}, be the case.)
Performing Monte Carlo simulations for $D=0$ and $D/R_\oplus=1.8$ at
$\eta>0.35$, two distinct structures, which are shown in Figs.~\ref{fig:2}(a)
and (b), can be observed.
The snapshots show an ordinary smectic-A phase for $D=0$ (a) and the phase
$S_{AW}$ for $D/R_\oplus=1.8$ (b).
While in panel (a) {one recognizes} a typical smectic layer structure with
{a} layer spacing comparable to the cation length $L_\oplus$, in panel (b)
{alternating} layers {are} observed, in which the cations (illustrated
in red) are well-aligned with the layer normal $\vec{\hat n}$, {as well as}
intermediate layers of cations (depicted in blue) {which are} oriented
almost perpendicularly to the layer normal.
Interestingly, the intra-layer structure is also different for the two phases.
For example for the two layers of the {common} phase $S_A$ and the
phase $S_{AW}$ in Figs.~\ref{fig:2} (c) and (d), {respectively}, one
observes a (typical) fluid-like structure for the phase $S_A$, while a dense
and fairly ordered structure is observed for the main layers of the phase
$S_{AW}$.

In order to discuss the structure of the two phases in more detail, the pair
distribution functions in the direction of the layer normal $\vec{\hat n}$,
i.e., $g_{ij}^{||}(n)$, and in lateral directions perpendicular to
$\vec{\hat n}$, i.e., $g_{\oplus\oplus}^{(n)}(r_\perp)$, (compare
Eqs.~\eqref{eq:PairDistFunc_normal} and \eqref{eq:PairDistFunc_lateral}) are
analyzed in Fig.~\ref{fig:3}.
In Figs.~\ref{fig:3}(a) and (i) the pair distribution functions are shown for
the phase $S_A$, as depicted in Fig.~\ref{fig:2}(a), for $D=0$.
{According to} the {red} curve in Fig.~\ref{fig:3}(a), the
cation-cation correlations, i.e., $g_{\oplus\oplus}^{||}(n)$, clearly show that
a layer structure with layer spacing $d/R_\oplus\approx3.5$, comparable to the
particle length $L_\oplus=4R_\oplus$, is formed.
Panel (i) shows, for {this} smectic-A structure, the lateral correlations
among the cations, i.e., $g_{\oplus\oplus}^{(n)}(r_\perp)$.
Within the layer in which the reference cation is located (black curve,
corresponding to $n=0$), clearly a fluid-like structure with rapidly decaying
correlations is observed.
For neighboring layers (magenta curve, i.e., $n=d$) one finds no
correlations at all.
Thus, these findings confirm that the structure shown in Fig.~\ref{fig:2}(a) is
an ordinary smectic-A phase ($S_A$).

The phase $S_{AW}$, formed for $D/R_\oplus=1.8$, is shown in Figs.~\ref{fig:3}
(c) and (iii).
The cation-cation correlations indicate the alternating layer structure
consisting of main layers of {high} cation density (the peaks of the
{red} curve at $n/R_\oplus=0$ and $n/R_\oplus\approx6$) and secondary
layers (at $n/R_\oplus\approx3$ and $n/R_\oplus\approx9$).
Interestingly, analyzing the lateral structure of the phase $S_{AW}$ in panel
(iii), we find a pronounced structure which is distinct from the (fluid-like)
pair distribution function obtained for the ordinary phase $S_A$.
The peak positions yield that this resembles a hexagonal structure, which is
also confirmed by the snapshot of a $S_{AW}$ layer, shown in Fig.~\ref{fig:2}
(d).
However, the lateral correlations within the neighboring layer with respect to
the reference cation (magenta curve in Fig.~\ref{fig:3}(iii)) are almost
vanishing and therefore the phase $S_{AW}$ is {indeed} a smectic phase and
not a crystal-like structure.
Thus, neighboring smectic layers can be sheared without any cost of free energy.
We note, that the {weak} oscillations, {which} are visible in the
magenta curve in Fig.~\ref{fig:3}(iii), are artifacts of the
periodic boundary conditions: If the layer normal is not parallel to one of the
main axis of the cubic simulation box, i.e., $\vec{\hat n}\notin\{\vec{\hat x},
\vec{\hat y},\vec{\hat z}\}$, {the smectic layers are not correctly
continued by periodic images of the simulation box.
For example, the smectic layer in the lower right corner of Fig.~\ref{fig:2}(b)
is continued to below by the periodic image of the third smectic layer counted
from the lower right corner.}
These artificial correlations (in lateral directions) between neighboring
smectic layers occur in principle also for the ordinary phase $S_A$.
{However}, due to the {short-ranged} lateral correlations, they are
not {visible} in Fig.~\ref{fig:3}(i).

{Presumably the} different structures of the phases $S_A$ and $S_{AW}$ are
directly related to the {slightly} {higher} density within the main
layers of the phase $S_{AW}$ as compared to the {layers of the} phase $S_A$
({see} the values of the cation-cation pair distribution function
{$g^\|_{\oplus\oplus}(n)$, i.e., the red curves in Figs.~\ref{fig:3}(a) and
(c), at $n=0$}).
The {higher} local cation densities within the $S_{AW}$ main layers are
stabilized by the Gay-Berne attraction for parallel oriented cations.
Yet, the charges at the tips are indispensable for the formation of the
phase $S_{AW}$ as they provide a net repulsion of neighboring smectic layers
and therefore make {it} energetically favorable to maintain a larger
distance between the dense main layers separated by the intermediate secondary
layers.
Given the relatively weak electrostatic interaction {as} compared to the
Gay-Berne attraction (see Fig.~\ref{fig:1}), already small density differences
within the (main) layers of the smectic-A phases decide on the stability of
$S_A$ or $S_{AW}$.
The large layer spacing in combination with the {slightly increased}
density in the main layers and the {substantially} lower density in the
secondary layers rationalizes the intermediate {mean} density {range}
in which the phase $S_{AW}$ is observed.

For $D=0$, however, the repulsion {between cations in} neighboring layers
is weaker, as compared to the case $D/R_\oplus=1.8$ ({see} the interaction
landscape for the two cases in Figs.~\ref{fig:1}(c) and (d)).
Thus, the energetic benefit of a larger layer spacing is {insufficient for
stabilizing} the phase $S_{AW}$ in favor of the phase $S_A$ with {the}
layer spacing {being} comparable to the cation length.

The comparison of the two cases, in which the cation charges are either
localized in the center, i.e., $D=0$, or the charges are {positioned}
close to the tips, i.e., $D/R_\oplus=1.8$, {underscores} the importance of
the charge distribution {within the cations} for the formation of the
phase $S_{AW}$: In agreement with previous results~\cite{Bartsch2017}, obtained
within the effective one-species model, one can conclude, that, due to the
cation charges at the tips, a considerable net repulsion of adjacent (main)
layers occurs at {small} distances (see Fig.~\ref{fig:1}(d)), which drives
the main layers apart to distances larger than the cation length $L_\oplus$. 

{In the case $D/R_\oplus = 1.8$, apparently} the incorporation of explicit
anions does not affect the formation of the phase $S_{AW}$.
However, as we {shall} present in the next {subsection}, the
distribution of the anions does sensitively depend on the charge distribution
within the cations.


\subsubsection{\label{sec:results:struc_dep:anions}Anion distribution}

The different cation charge distributions for $D=0$ and $D/R_\oplus=1.8$
not only lead to {remarkably different smectic} structures, formed by the
ellipsoidal cations, {but} moreover, for the two cases the distribution of
anions is also distinct.

Revisiting Fig.~\ref{fig:3}(a), which depicts the pair distribution functions
$g_{ij}^{||}(n)$ along the layer normal for the ordinary smectic-A phase $S_A$
for $D=0$, one finds that the anions are rather homogeneously distributed
around the layers of cations.
({See} the {green and blue} curves {which} show only minor
{spatial} variations.)
{The pair distribution function} $g_{\oplus\ominus}^{||}(n)$ (blue curve)
shows a sparse tendency of the anions to be located in between the cation
layers.
In contrast, for the phase $S_{AW}$ at the same temperature $T^*=0.55$ the
anions are strongly pushed out of the main layers {formed} by the cations.
The highest probability to find the anions is close to the location of the
charges of the cations in the main layers, i.e., the maxima of
$g_{\oplus\ominus}^{||}(n)$ are found {at} a distance $n/R_\oplus\approx
D/R_\oplus=1.8$ away from the centers of the smectic layers. 

A {similarly} strong inhomogeneous distribution of anions is found for the
ordinary phase $S_A$ which is formed by cations with $D/R_\oplus=1.8$
{and} at higher temperatures.
In Fig.~\ref{fig:3}(d) the phase $S_A$, for $D/R_\oplus=1.8$ at {the}
slightly higher temperature $T^*=0.58$, is analyzed.
While the structure (in normal direction) of the cations ({red} curve
{in panel (d)}) is very similar to the phase $S_A$ formed for $D=0$ (see
{Figs.~\ref{fig:3}(a) and (i)}), analogously to the case of the phase
$S_{AW}$, the anions are preferentially located in between the smectic layers
formed by the cations (see the blue curve in panel (d)).

The same findings are obtained for the respective crystal-like phases
{which} are observed in both cases, i.e., for $D=0$ and $D/R_\oplus=1.8$,
at low temperatures and for large densities (see the detailed discussion of the
phase {behavior} for the two cases in the {following
Subsec.}~\ref{sec:results:Phases}).
Thus we conclude that, for charges at the tips of the cations, not only layer
structures of the cations are observed, but also for the anions, although the
inhomogeneity in the distribution of {the} anions is not as pronounced as
for the cations.
{In line with these findings}, for $D=0$, i.e., if the charges of the
cations are localized in the centers, layering is only observed for the
cations.
Supposedly, this structural behavior is driven by the interplay of, {on
the one hand}, the electrostatic attraction of the anions towards the cation
centers and, on the other hand, {with} the steric hard-core repulsion,
which hinders the anions to penetrate the cation layers.

This striking difference in the anion distributions for the two cases is also
relevant for potential applications of ILCs, because some technologies
incorporating ILCs like \emph{dye-sensitized solar cells} (DSSCs) benefit from
a higher density of counterions, as they are needed as charge {carriers}
in these applications~\cite{Yamanaka2005}.
Thus, ILCs with charges at the tips, seem to be the best {candidates} for
such applications, {whereas cations with} $D=0$ {seem} to be good
{candidates} if a smectic phase of cations in combination with a
homogeneous distribution of anions is required.


\subsection{\label{sec:results:Phases}Phase behavior}

\begin{figure}
\includegraphics[width=8cm]{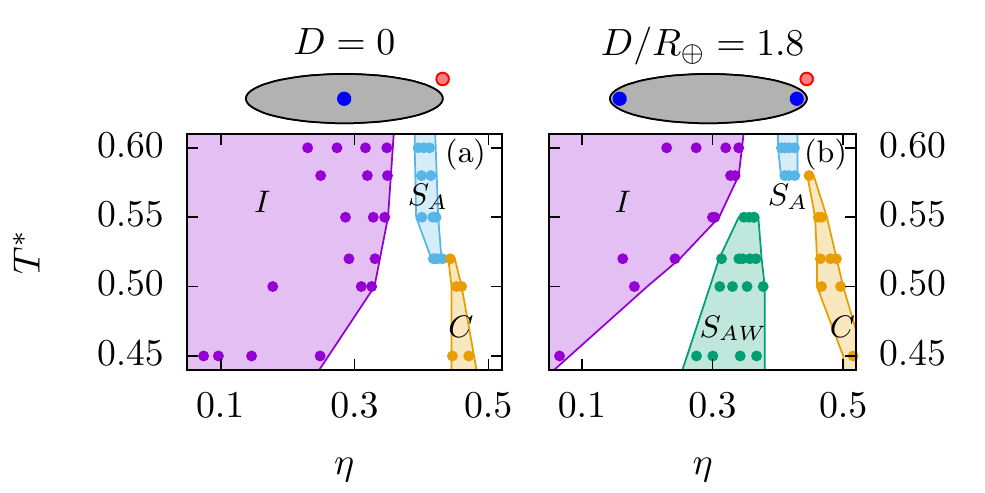}
\caption{The panels show the phase diagrams for (a) $D=0$ and (b)
 $D/ R_\oplus=1.8$.
 The colored areas represent estimates of the \emph{one}-phase regions which
 are obtained by determining thermodynamically stable states (colored dots)
 generated by the grand canonical MC simulations with isotropic
 initializations.
 The white areas are estimates of the \emph{two}-phase regions, the extensions
 of which are resolved with an accuracy of $\Delta\eta\leq0.01$ in terms
 of the packing fraction $\eta$ (for details see the beginning of
 Sec.~\ref{sec:results}).
 It has been checked by means of short additional simulation runs with
 smectic-A and crystalline initial configurations that the results of the
 simulations do not depend on the initialization.
 For $D=0$ (a) at high temperatures $T^*\gtrsim0.52$ a first-order phase
 transition {occurs} from the isotropic fluid phase $I$ (violet-colored area) to the
 ordinary smectic-A phase $S_A$ (blue), while at
 lower temperatures a direct first-order phase transition {takes place to} the
 crystalline phase $C$ (yellow) with a hexagonal lattice structure.
 In contrast, for $D/R_\oplus=1.8$ (b) at low temperatures the phase $S_{AW}$
 (green) is stable {within} an intermediate density {range}, such
 that, upon increasing $\eta$, first one observes a discontinuous phase
 transition from the isotropic fluid $I$ to the phase $S_{AW}$, followed at
 large densities by a first-order phase transition from the phase $S_{AW}$ to
 the hexagonal crystal $C$.}
\label{fig:4}
\end{figure}

{After having} analyzed the differences {of} the smectic structures
{associated with} the charge distributions $D=0$ and $D/R_\oplus=1.8$, we
now {focus} on the phase behavior, i.e., {identifying those regions in
the $(T^*,\eta)$ plane, which correspond to the aforementioned distinct
structures.}

First, we {analyze} the phase behavior of ILCs consisting of cations with
{the} charges {localized in their center}, i.e., {for} $D=0$.
{Figure~\ref{fig:4}(a) shows} the $T^*$-$\eta$ phase diagram for $D=0$.
{Within} the investigated temperature range $T^*\in[0.45,0.6]$, the
isotropic fluid phase $I$ ({violet}) is {thermodynamically} stable up to
(cation) packing fractions {$\eta\approx0.25\ {\dots}\ 0.35$.}
At sufficiently high temperatures $T^*\geq0.52$ a first-order phase transition,
{indicated by a density gap}, is observed to the ordinary smectic-A
phase $S_A$ (blue).
However, at lower temperatures, a direct phase transition {to} a
crystal-like structure $C$ ({yellow}) {occurs}.
While the phase $S_A$ and the crystalline phase $C$ show a rather similar layer
spacing $d/R_\oplus\approx3.5$ in normal direction (compare the pair
distribution functions $g_{\oplus\oplus}^{||}(n)$ shown in Figs.~\ref{fig:3}(a)
and (b)), {their} lateral structures are quite distinct.
For the phase $C$ a hexagonal ordering (see Fig.~\ref{fig:3}(ii)) can be
observed and, moreover, {there are} strong correlations between
neighboring layers.
Indeed, comparing the observed peaks {in the distribution functions} with
an ideal {three-dimensional} hexagonal lattice, one finds very good
agreement concerning the peak {positions} (in the $0$-th layer, as well
as in the neighboring layers).
Interestingly, although {in general} the anions {have} a tendency to
be {localized} in between the cation layers in the phase $C$, {here}
the distribution of anions is fairly homogeneous ({relative} to the
pronounced peaks in the cation distribution).
Thus, while at the considered high densities the cations form a well-marked
hexagonal lattice, the anions are still rather motile.

Now {we turn} to the ILCs with {the} cation charges at the tips,
i.e., $D/R_\oplus=1.8$.
{The corresponding} phase diagram is shown in Fig.~\ref{fig:4}(b).
{At high temperatures $T^*\gtrsim0.58$} one also {finds} a
first-order phase transition from the isotropic fluid $I$ {to} the
ordinary smectic phase $S_A$.
Besides the additional inhomogeneous distribution of {the} anions, which
has not been observed for the phase $S_A$ for $D=0$ (see
Sec.~\ref{sec:results:struc_dep:anions}), {here the} smectic-A phase is
very similar to {the} phase $S_A$ {forming} for $D=0$.
The layer spacing $d/R_\oplus\approx3.5$ is comparable to the cation length
$L_\oplus$ and the cations are well-aligned with the layer normal.
Furthermore, the lateral correlations (see Fig.~\ref{fig:3}(iv)) clearly
{exhibit} a fluid-like structure within the smectic layers.
If the temperature is lowered to {$T^*\lesssim0.55$}, a different
structure {appears}.
From Figs.~\ref{fig:3}(c) and (iii) one infers that this distinct structure
corresponds to the phase $S_{AW}$ (green).
{Since} the phase $S_{AW}$ emerges in an intermediate density region,
for $D/R_\oplus=1.8$ the isotropic fluid $I$ is stable only at small
densities at low temperatures.   
An alternating structure of main and secondary layers occurs which leads to a
significantly increased layer spacing $d/R_\oplus\approx6.0$.
Due to {this} increased layer spacing, driven by the electrostatic
repulsion of neighboring main layers, the bulk densities {$\eta$} of the
phase $S_{AW}$ are {lower} than the bulk densities {$\eta$} at which
the ordinary phase $S_A$ is observed.
However, by further increasing the number of particles in the system (via
raising the chemical potential), at sufficiently {high} densities the
phase $S_{AW}$ becomes metastable with respect to the hexagonal crystal $C$
(see Fig.~\ref{fig:4}(b) for {$T^*\lesssim0.55$ and $\eta\gtrsim
0.4$}).

Comparing the present findings with the theoretical predictions of
Ref.~\cite{Bartsch2017} ({see} Fig.~5 therein), it is remarkable that the
DFT results for the effective one-species model, on a qualitative {level},
predict the stability of the phase $S_{AW}$ in the same thermodynamic region as
{the} present Monte Carlo simulations {for} the enhanced ILC model, i.e.,
at low temperatures and at intermediate densities.
Furthermore, {in both approaches} the charge distribution of the
(ellipsoidal) cations {turns out to be} crucial for the formation of the
phase $S_{AW}$, i.e., {in order to have a stable phase $S_{AW}$} it is
{indispensable} that the charges are close to the tips of the cations.

Finally, it is worth mentioning, that although no stable state points of the
phase $C$ {have been found} at high temperatures, {i.e.,
$T^*\gtrsim0.52$} for $D=0$ {and $T^*\gtrsim0.58$} for $D/R_\oplus=
1.8$, {respectively}, at sufficiently large densities a crystal-like phase
{is expected to} be the stable configuration, {even} at high temperatures.
Supposedly, with our MC simulations we {have been unable} to reach the
very high density {region, which requires} a large number of particles in
the simulation box {and thus slows down the simulations}.


\section{\label{sec:conclusions}Summary and conclusions}

The objective of the present study is to shed light on the role of counterions
in {forming} smectic structures in ionic liquid crystals (ILCs).
In particular, our {analysis} aims {at} investigating how the phase
{behavior} and {the} structural {properties} of {an
effective} one-species model, {which has been employed in previous}
theoretical studies {of} ILCs~\cite{Kondrat2010,Bartsch2017,Bartsch2019},
{are} affected by {taking the} counterions explicitly {into
account}.
{These} previous {models represent} a simplified description of ILCs,
{which are} composed of anisotropic mesogenic ions (for typical examples,
these are cations~\cite{Binnemans2005,Goossens2016}), which are embedded in a
homogeneous screening background {consisting} of much smaller anions. 

The present model, {which incorporates both} cations and anions {on
equal footing}, can be understood as an {improvement} of the previous one,
because it does not {take at the outset} any specific distribution of the
anions.
Thereby it allows {one} to investigate not only the {influence} of
the anions on the previously observed liquid-crystalline
structures~\cite{Bartsch2017}, {but}, moreover, the anion distribution by
itself can be analyzed, which is of particular interest for potential
technological applications of ILCs, e.g., as electrolyte materials in solar
cells~\cite{Yamanaka2005}.

The current coarse-grained model (see Sec.~\ref{sec:model} and
Fig.~\ref{fig:1}) exhibits rigid ellipsoidal cations with two charge sites,
symmetrically located at a distance $D$ from the molecular center and spherical
anions with one central charge site.
All results of this work have been obtained using {grand canonical} Monte
Carlo (MC) simulations.

Depending on the intra-molecular cation charge distribution, distinct smectic
structures are observed.
They are formed upon increasing {the} density (expressed in terms of the
cation packing fraction $\eta$) such that the isotropic fluid $I$, which is the
stable phase at low densities, becomes metastable (and ultimately unstable)
with respect to {forming} smectic layer structures. 
For $T^*=0.55$ a first-order phase transition from the phase
{$I$ to} an ordinary smectic-A phase $S_A$ is observed for $\eta>0.35$, if
the cations carry a single charge site in {their} center, i.e., for $D=0$
(see Fig.~\ref{fig:2}(a)).
The designation of the {emerged} structure as 'smectic-A' is based on the
following observations.
The ellipsoidal cations form layers in which they mostly orient parallel to the
layer normal $\vec{\hat n}$ and the layer spacing $d\approx3.5R_\oplus$
(Fig.~\ref{fig:3}(a)) is of the size of the cation length $L_\oplus$.
Moreover, in {the} directions {which are lateral} with respect to
$\vec{\hat n}$, i.e., within the smectic layers, a fluid-like structure is
observed (see {Figs.}~\ref{fig:2}(c) and \ref{fig:3}(i)).

In contrast for $D/R_\oplus=1.8$, i.e., {for} cations with charges at
{their} tips, {at} the same temperature $T^*=0.55$ ({and} at
sufficiently {high} densities, i.e., {$\eta\gtrsim0.3$}) a layer
structure, {which is} distinct from the ordinary phase {$S_A$}, is
found: Alternating layers of cations {are observed, which are} mostly
parallel to the layer normal $\vec{\hat n}$, and of cations, {which are}
oriented mostly perpendicular to $\vec{\hat n}$ (Fig.~\ref{fig:2}(b)).
This structure can be identified as the wide smectic-A phase $S_{AW}$, which
has been {found previously}~\cite{Bartsch2017}.
In agreement with the previous findings, the (main) layers, in which the cations
are well aligned, show a much larger (local) density as compared
{with} the (secondary) layers, in which the cations are {mostly}
perpendicular to $\vec{\hat n}$ (see $g_{\oplus\oplus}^{||}(n)$ in
Fig.~\ref{fig:3}(c)).
Moreover, the layer spacing $d\approx6R_\oplus$ of the phase $S_{AW}$ is
significantly larger than for the ordinary smectic-A phase $S_A$.
Due to the {high} density in the main layers, a lateral hexagonal
structure is observed (see Figs.~\ref{fig:2}(d) and \ref{fig:3}(ii)).
However, {there are no visible} correlations among neighboring layers and
therefore the phase $S_{AW}$ is a genuine smectic phase and not a crystal.

Interestingly, {from} comparing the distribution of anions in normal
direction {$\vec{\hat n}$}, we {infer} that for $D=0$ the anions are
rather homogeneously distributed around the cation layers
({Figs.}~\ref{fig:3}(a) and (b)), while for $D/R_\oplus=1.8$ a pronounced
localization of anions in between the cation layers is observed.
While for $D=0$ the competing electrostatic attraction and the steric
repulsion of cations and anions for small center-to-center distances presumably
lead to the fairly homogeneous distribution of anions, for $D/R_\oplus=1.8$
the anions are not strongly inhibited by steric repulsion to accumulate at the
tips of the cations.

{Concerning} the phase behavior (see the phase diagrams in
Fig.~\ref{fig:4}) for the currently studied model, we find a remarkable
(qualitative) agreement with the previously studied one-species model
description of ILC systems.
The phase {$S_{AW}$} is formed only if the cation charges are
{positioned} at {their} tips.
Furthermore, its stable {region} is found at lower temperatures, as
compared to the ordinary smectic-A phase $S_A$, and at intermediate densities,
i.e., in particular at lower densities than the {ones for the} stable
phase {$S_A$} and at higher densities than the {ones of the} stable
isotropic fluid $I$, which is in agreement with the phase behavior predicted by
DFT~\cite{Bartsch2017}.
In both cases, i.e., for $D=0$ and $D/R_\oplus=1.8$, at very {high}
densities a three-dimensional hexagonal crystal $C$ is formed (see
{Figs.}~\ref{fig:3}(b) and (ii) {as well as Figs.}~\ref{fig:3}(e) and
(v), {respectively}).

These results are not only consistent with the previous findings for the
effective one-species description of ILCs, {but}, moreover, they pinpoint
the significance of the (intra-molecular) charge distribution for the phase 
behavior {as} well as for the structural {properties} of ILC systems.

Future studies might focus on the dependence of the thermal {behavior} and
structural {properties} of ILCs on the anion size and shape, but also on
the strength of the Gay-Berne potential {as compared to the electrostatic
interactions}.
This, in particular, is a subtle issue, because ILCs typically exhibit an
effective charge which is even less than one elementary charge, due to effects
like charge delocalization~\cite{Saielli2017}. 


\begin{acknowledgments}
We thank C.\ Holm for valuable comments.
\end{acknowledgments}

\end{document}